\documentclass[12pt,
preprint,aps,prc,showpacs,superscriptaddress,tightenlines]{revtex4}
\usepackage{epsf}
\usepackage{amsmath}

\def\beq{\begin{equation}}
\def\eeq{\end{equation}}
\def\bea{\begin{eqnarray}}
\def\eea{\end{eqnarray}}
\def\beqa{\begin{equation}\begin{array}{l}}
\def\eeqa{\end{array}\end{equation}}
\def\eqlab#1{\label{eq:#1}}
\def\figlab#1{\label{fig:#1}}

\def\eref#1{(\ref{eq:#1})}
\def\Eqref#1{Eq.~(\ref{eq:#1})}

\def\Figref#1{Fig.~\ref{fig:#1}}

\def\quarter{\mbox{\small{$\frac{1}{4}$}}}

\def\barr{\left(\begin{array}{c}}
\def\earr{\end{array}\right)}
\def\bmat{\left(\begin{array}{cc}}
\def\emat{\end{array}\right)}
\def\al{\alpha}
\def\be{\beta}
\def\ga{\gamma} 
\def\de{\delta} \def\De{\Delta}

\def\la{\lambda}

\def\si{\sigma} 
  
\def\w{\omega}

\def\pa{\partial}

\def\pa{\partial}

\def\nn{\nonumber}

\def\mathscr{\mathcal}

\def\3d{3-D}

\def\amm{a.m.m.}

\begin{document}
\preprint{WM-04-111}

\title{A Derivative of the Gerasimov-Drell-Hearn Sum Rule}

\author{Vladimir Pascalutsa}
\email{vlad@jlab.org}
\affiliation{Theory Group, JLab, 12000 Jefferson Ave, Newport News, VA 23606}
\affiliation{Department of Physics, College of William \& Mary, Williamsburg, VA
23188}

\author{Barry R.\ Holstein}
\email{holstein@physics.umas.edu} \affiliation{Theory Group, JLab,
12000 Jefferson Ave, Newport News, VA 23606}
\affiliation{Department of Physics--LGRT, University of
Massachusetts, Amherst, MA 01003}

\author{Marc Vanderhaeghen}
\email{marcvdh@jlab.org}
\affiliation{Theory Group, JLab, 12000 Jefferson Ave, Newport News, VA 23606}
\affiliation{Department of Physics, College of William \& Mary, Williamsburg, VA
23188}
\date{\today}

\begin{abstract}
We derive a sum rule which establishes a linear relation between a
particle's anomalous magnetic moment and a quantity connected to
the photoabsorption cross-section. This quantity cannot be
measured directly.  However, it can be computed within a given
theory.  As an example, we demonstrate validity of the sum rule in
QED at tree level---the renowned Schwinger's correction to the
anomalous magnetic moment is readily reproduced.  In the case of
the strong interactions, we also consider the calculation of the
nucleon magnetic moment within chiral theories.
\end{abstract}

\pacs{11.55.Hx, 13.40.Em, 25.20.Dc}%
\maketitle
\thispagestyle{empty}


The well-known Gerasimov-Drell-Hearn (GDH) sum rule
(SR)~\cite{GDH}, \beq \eqlab{gdhsr} \frac{\pi \alpha }{M^2 s}
\,\kappa^2 = \frac{1}{\pi}\int\limits_{0}^\infty \!
\frac{d\w}{\w}\,\De\si^{(\mathrm{full})}(\w), \eeq
relates the anomalous magnetic
moment (\amm)\footnote{Here by the anomalous magnetic moment we understand
$\kappa=(g-2)s$, i.e., the deviation of the gyromagnetic ratio $g$ from
its natural value of 2 for any spin\cite{fpt}.} $\kappa$ of a particle with spin $s$ and mass
$M$ to the integral of the difference of
polarized total photoabsorption cross-sections:
\beq
\De\si^{(\mathrm{full})}_s(\w)=\si^{(\mathrm{full})}_{1+s}(\w)
- \si^{(\mathrm{full})}_{1-s}(\w)\, ,
\eeq
where
$\sigma_{1 \pm s}^{(\mathrm{full})}$ are the photoabsorption cross sections for
total helicity $(1 \pm s)$, and the superscript `full' refers to the sum over
all possible final states.  Below we consider only the case of a
spin-1/2 particle, hence $\De\si=\si_{3/2} - \si_{1/2} $.
Furthermore, in \Eqref{gdhsr}, $\omega$ is the photon {\it
laboratory} energy in the photoabsorption reaction, while $\alpha
= e^2 / 4 \pi \simeq 1 / 137$ is the fine structure constant.

The derivation of this SR is based on the general principles of
{\it analyticity} (dispersion theory, with the assumed validity of
an unsubtracted dispersion relation), {\it unitarity} (optical
theorem), electromagnetic {\it gauge-invariance}, and {\it
crossing symmetry}. Therefore any theory that satisfies these
fundamental principles and which has the appropriate convergence
property should be consistent with this SR, in the sense that the
left-hand-side and the right-hand-side of \Eqref{gdhsr}, computed
within that theory, must agree.

In QED, for instance,  this consistency has long ago been
verified, to lowest order in $\alpha$.  Since in this case the
lowest order contribution to the \amm\ is of order $\alpha$, the
{\it lhs} of the GDH SR \Eqref{gdhsr} starts at ${\cal O}(
\alpha^3)$. On the {\it rhs}, however, the tree-level contribution
to the cross section difference $\De\si$ is non-vanishing at order
$\alpha^2$. Hence, in order to satisfy the GDH SR at order
$\alpha^2$ the lowest order integral in \Eqref{gdhsr} must
vanish~: \beq 0=\int\limits_{0}^\infty \! \frac{d
\w}{\w}\,\,\De\si^{(\mathrm{tree})}(\w)\,,\label{eq:no} \eeq which
requires a careful cancellation between low and high frequency
components of the photoabsorption cross sections. That this
cancellation occurs precisely in QED and more generally
in the standard electroweak theory has been verified by
Altarelli, Cabibbo, and Maiani~\cite{wk}.

In order to check the consistency of the GDH SR at ${\cal O}(
\alpha^3)$ in QED, which corresponds with the familiar Schwinger
correction to the \amm\ --- $\alpha/2\pi$ --- one needs to consider all
one-loop electromagnetic corrections to the photoabsorption cross
sections on the {\it rhs} of \Eqref{gdhsr}.   At this order, using
the calculations of Tsai, deRaad, and Milton\cite{tdm} the SR has
only recently been verified by Dicus and Vega~\cite{DiV01}. This
calculation turns out to be rather involved, since one needs to
compute the one-loop Compton amplitude, plus consider other
inelastic processes, such as pair creation, then in the end
perform the integral over photon energy.  This allows the recovery
of Schwinger's result which, of course, can be obtained much more
easily by computing the one-loop vertex correction\cite{bjd}.
Thus, although one can obtain the same result for the \amm\ by
computing either the usual vertex correction or via the {\it rhs}
of the GDH SR, the direct one-loop calculation is clearly much
more straightforward and economical.

In this note we shall present a new sum rule, for which the
calculation of the dispersion integral over the photoabsorption
cross section turns out to be {\it simpler} than the direct loop
calculation---Schwinger's result can be obtained by integrating a
{\it tree-level} cross section.

This sum rule can be derived from the GDH SR by introducing a
``classical'' value of the \amm, $\kappa_0$. At the
field-theoretic level this implies that Dirac's Lagrangian of the
spin-1/2 field acquires an explicit Pauli term: \beq {\cal
L}_{Pauli} = i\frac{e\kappa_0}{4M}\,\bar\psi\, \sigma^{\mu\nu}
\,\psi \,F_{\mu\nu}\,, \eeq where $\sigma^{\mu\nu}={i\over 2}
[\ga^\mu,\ga^\nu]$ and $F_{\mu\nu}=\partial_\mu A_\nu-\partial_\nu
A_\mu$ is the electromagnetic field tensor.  The total value of
the \amm\ is then \beq \eqlab{total} \kappa = \kappa_0 + \de
\kappa, \eeq where by $\de \kappa$ we denote all quantum (loop)
corrections.

At this stage it is important to realize that in this theory with
an explicit Pauli term the GDH SR is not valid, since there now
exists a tree-level contribution to the Compton scattering
amplitude which cannot be reproduced by a dispersion relation
using the degrees of freedom included in the theory ({\it e.g.},
photons and spin-1/2 fermions in case of QED). In fact, this
tree-level contribution to the photoabsorption cross section, proportional to
$\kappa_0^2$, corresponds with the d.o.f. which are integrated out
of the theory, and can be represented as~:
\beq
\eqlab{gdhk0}
\frac{2 \pi \alpha}{M^2}\, \kappa_0^2 =
\frac{1}{\pi}\int\limits_{0}^\infty \! \frac{d\w}{\w}\,\De\si^{(\mathrm{he})}(\w),
\eeq
where $\Delta \sigma^{(\mathrm{he})}$ is the
photoabsorption cross sections with the `high energy' d.o.f.
integrated out of the theory. By subtracting \Eqref{gdhk0} from
\Eqref{gdhsr}, and expanding $\kappa$ as in \Eqref{total}, we obtain the
sum rule~:
\beq
\eqlab{gdhsr3}
\frac{2 \pi \alpha}{M^2}\,
\left\{ (\delta \kappa)^2 + 2\,\kappa_0\, \delta \kappa \right\} =
\frac{1}{\pi}\int\limits_{0}^\infty \! \frac{d\w}{\w}\, \De\si(\w; \kappa_0),
\eeq
where $\De\si \equiv \De\si^{(\mathrm{full})} - \De\si^{(\mathrm{he})}$
corresponds to the photoabsorption cross-section that involves the d.o.f.
included in the theory. Furthermore, we have indicated that the
photoabsorption cross section depends explicitly on $\kappa_0$.
Note that also $\de\kappa$ will in general depend on $\kappa_0$, as the Pauli coupling
appears in the loops.

By taking the limit to the theory with {\it vanishing}
classical \amm, {\it i.e.} $\kappa_0=0$, one has $\delta \kappa
\rightarrow \kappa $ and therefore \Eqref{gdhsr3} is nothing else
than the GDH SR.
We can however obtain a new sum rule by taking the
first derivative with respect to $\kappa_0$ of both {\it lhs} and {\it rhs} of
\Eqref{gdhsr3} and then let $\kappa_0 = 0$, which yields~:
\beq
\eqlab{linearsr}
\frac{4 \pi \alpha}{M^2}\, \kappa \,
\left( 1 + \left[ \frac{\partial}{\partial \kappa_0} \delta \kappa
\right]_{\kappa_0 = 0} \right) =
\frac{1}{\pi}\int\limits_{0}^\infty \! \frac{d\w}{\w}\, \De\si_1(\w),
\eeq
where we introduced the definition
\beq
\Delta\sigma_1(\omega)=\left[{\partial \Delta\sigma(\omega,\kappa_0)\over
\partial \kappa_0}\right]_{\kappa_0=0},
\eeq corresponding to the first derivative of the physical cross
section with respect to $\kappa_0$. Note that in the derivation of Eq. (8) we did 
{\it not} rely on any perturbative expansion
in coupling constants. This sum rule should then be valid non-pertubativelu as well as to any given order in perturbation theory. To lowest order in perturbation theory
the term in the brackets on the left-hand side of  \Eqref{linearsr} reduces
to unity, yielding~: \beq \eqlab{linearsr2} \frac{4 \pi
\alpha}{M^2}\, \kappa \, =\, \frac{1}{\pi}\int\limits_{0}^\infty
\! \frac{d\w}{\w}\, \De\si_1(\w) .
\eeq
The striking feature of this sum rule is the {\it linear}
relation between the \amm\ and the photoabsorption cross section,
in contrast to the GDH SR where the relation is quadratic.  We wish to point out that,
certainly, the cross-section quantity $\De\si_1$ is not an
observable in the theory with $\kappa_0=0$. However, it is very
clear how to determine it within a given theory.

As an example we consider QED, wherein to lowest order in $\al$
the photoabsorption cross-section is given by the Compton
scattering tree graphs, see \Figref{compton_qed}. Both the cross
section and its $\kappa_0$ derivative at $\kappa_0=0$ can be straightforwardly
computed to order $\alpha^2$ with the result: \bea
\De\sigma(\omega)& =& \frac{2\pi\alpha^2}{M^2x}\left[2+{2x^2\over
(1+2x)^2} - \left(1+{1\over
x}\right)\ln(1+2x)  \right]\label{eq:gdh} , \\
\De\sigma_1(\omega)& =& \frac{2\pi\alpha^2}{M^2x}\left[ 6-{2x
\over (1+2x)^2} -\left(2+{3\over x}\right)\ln(1+2x) \right] ,
\label{eq:gdh2} \eea where  $x=\omega/M$.  As mentioned above,
consistency of the GDH sum rule requires that the GDH integral of
$\De\sigma$ vanishes\cite{wk}---{\it cf.} \Eqref{no}---and
this is easily verified for the form obtained in \Eqref{gdh}. As
seen from \Figref{qedfig}, where we plot the corresponding
integrand, there exists an exact cancellation between the low- and
high-energy pieces of the cross section.

Evaluating the same type of integral for the derivative $\De\sigma_1$,
shown by the dashed curve in \Figref{qedfig}, we find~:
\begin{equation}
\frac{1}{\pi}\int\limits_0^\infty \!{d\omega\over \omega }\, \De\sigma_1 (\w)=
\frac{2\al^2}{M^2} \,.
\end{equation}
Thus, employing  the linearized GDH sum rule \Eqref{linearsr2} we obtain
$\kappa=\al/2\pi$---Schwinger's value!  We wish to emphasize that
this result has been obtained by calculating a {\it tree-level}
cross section derivative with respect to the a.m.m., {\it i.e.},
\Eqref{gdh2}, and then performing the dispersion integral. That we
obtain the same result as the usual calculation of a one-loop
integral is a direct consequence of the general principles of
analyticity, unitarity, gauge invariance, crossing symmetry, and
convergence leading to an unsubtracted dispersion integral.

As a further example of how dispersion relations provide an
economical method to evaluate one-loop corrections by calculating
only tree level cross sections, we consider the theory of nucleons
interacting with pions via pseudovector coupling: \beq {\cal
L}_{\pi NN} = \frac{g}{2 M}\, \bar\psi \,\ga^\mu \,\ga^5 \,\tau^a
\,\psi \,\pa_\mu \pi^a, \eeq where $g$ is the pion-nucleon
coupling constant, $\tau^a$ are isospin Pauli matrices, $\pi^a$ is
the isovector pion field. To lowest order in  $g$ the
photoabsorption cross section in this theory is dominated by the
single pion photoproduction graphs as displayed in \Figref{Born_chpt}.
We find for the corresponding GDH cross sections:
\begin{subequations}
\eqlab{piprod}
\bea
\De\si^{(\pi^0 p)} &=&
\frac{\pi C}{M^2 x^2}\,\left[  (2\al \bar s+1-x) \ln\frac{\al+\la}{\al-\la} -
2 \la[ x (\al-2) + \bar s (\al+2) ] \right], \qquad \\
\De\si^{(\pi^+ n)} &=& \frac{2\pi C}{M^2 x^2}\,
\left[  -\mu^2 \ln\frac{\be+\la}{\be-\la} + 2\la(\bar s\be-x\al)\right], \\
\De\si^{(\pi^0 n)} &=& 0,\\
\De\si^{(\pi^- p)} &=&
\frac{2\pi C}{M^2 x^2}\,\left[ -\mu^2 \ln\frac{\be+\la}{\be-\la} + (2\al \bar
s-1-x) \ln\frac{\al+\la}{\al-\la} -
2\bar s\la\right],
\eea
\end{subequations}
where $ C=\left(eg/4\pi\right)^2$, $\mu = m_\pi/M $, $m_\pi$ is
the pion mass, and
\begin{subequations}
\bea
&& s= M^2 + 2M\w,\,\,\, \bar s = s/M^2,\\
&& \al=(s+M^2-m_\pi^2)/2s,\\
&& \be =(s-M^2+m_\pi^2)/2s=1-\al ,\\
&& \la = (1/2s)\sqrt{s-(M+m_\pi)^2}\sqrt{s-(M-m_\pi)^2} \,.
\eea
\end{subequations}

As in the case of QED, the anomalous magnetic moment corrections
start at ${\cal O}(g^2)$, implying that the $lhs$ of the GDH
SR begins at ${\cal O}(g^4)$.  Since the tree-level cross
sections are ${\cal O}(g^2)$, we must require that \beq
\eqlab{conc1} \int\limits_{\w_{\rm th}}^\infty {d\omega\over
\omega }\, \De\sigma^{(I)} (\w) =0, \,\,\,\, \mbox{for $I=\pi^0
p,\,\pi^+ n,\, \pi^0 n, \pi^- p$}, \eeq where $\w_{\rm th} = m_\pi
(1+ m_\pi/2M)$ is the threshold of the pion photoproduction
reaction. This requirement is indeed verified for the expressions given in \Eqref{piprod} ---the consistency of GDH SR is maintained in this theory for each of the pion production
channels.

Again, as in QED, there exists an intricate cancellation between
the low- and high-energy component of these cross sections, {\it
cf.} \Figref{chpt1} (solid curves).  Any attempted approximation
to these cross-section, {\it e.g.}, by means of semi-relativistic
or chiral expansions could in principle violate this consistency
with the GDH SR.  This is indeed the case for the non-relativistic
CGLN expressions~\cite{CGLN}, which are known to violate the
SR~\cite{Lvov93}.  It is plausible, however, that any such
truncations can be performed in a fashion that maintains the
consistency. For example, for the $\pi^+$ production the CGLN
expression can be supplemented by a term which is formally
higher-order in $\w/M$ to yield a cross-section: \beq
\De\si^{(\pi^+ n)}=\frac{2\pi C}{M^2}\left[-\frac{m_\pi^2}{\w^2}
\, \ln
\left(\frac{\sqrt{\w^2+m_\pi^2}+\w}{\sqrt{\w^2+m_\pi^2}-\w}\right)
+\frac{2\w}{M\bar s^2} \right] ,
\eeq which still satisfies \Eqref{conc1}.
It is interesting to observe that the addition of this new term
(the second term in the square brackets) allows the
semirelativistic approximation to have the same chiral limit as
the full relativistic result: \beq \lim_{m_\pi\rightarrow 0}
\De\si^{(\pi^+ n)}  = \frac{2\pi C}{M^2}
\left[-\w\,\de(\w)+\frac{2\w}{M\bar s^2} \right]\,. \eeq  and from
this expression it is particularly easy to see that the
consistency is maintained in the chiral limit: $\int_0^\infty
\!{d\omega\over \omega}\lim\limits_{m_\pi\rightarrow 0}
\De\si^{(I)}=0$.

We now turn our attention to the linearized GDH sum rule. In this
case we first introduce Pauli moments $\kappa_{0p}$ and
$\kappa_{0n}$ for the proton and the neutron, respectively. The
dependence of the cross-sections on these quantities can generally
be presented as: \bea \De\si(\w;\, \kappa_{0p},\, \kappa_{0n})
&=&\De\si(\w)
+\kappa_{0p} \,\De\si_{1p}(\w) + \kappa_{0n}\, \De\si_{1n}(\w) \nn\\
&+& \kappa_{0p}^2 \,\De\si_{2p}(\w) + \kappa_{0n}^2\,
\De\si_{2n}(\w) + \kappa_{0p}\,  \kappa_{0n}\, \De\si_{1p1n}(\w) +
\ldots. \eea Furthermore, we introduce proton and neutron
photoproduction cross sections  $\De\si^{(p)}$ and $\De\si^{(n)}$
and express the corresponding GDH SRs, corrected by the
$\kappa_0^2$ terms.  Analogous to the QED case, we obtain~:
\begin{itemize}
\item[(i)]  the GDH SRs:
\begin{subequations}
\eqlab{newsrs}
\beq
\frac{2 \pi \alpha}{M^2}\,\kappa_p^2 =\frac{1}{\pi}\int\limits_{\w_{th}}^\infty \!
\frac{d\w}{\w}\,\De\si^{(p)},\,\,\,\,
\frac{2 \pi \alpha}{M^2}\,\kappa_n^2 =\frac{1}{\pi}\int\limits_{\w_{th}}^\infty \!
\frac{d\w}{\w}\,\De\si^{(n)},
\eeq
\item[(ii)]  the {\it linearized} SRs
(valid to leading order in the coupling $g$): \beq \frac{4 \pi
\alpha}{M^2}\, \kappa_p =
\frac{1}{\pi}\int\limits_{\w_{th}}^\infty \!
\frac{d\w}{\w}\,\De\si_{1p}^{(p)} ,\,\,\,\, \frac{4 \pi
\alpha}{M^2}\, \kappa_n =
\frac{1}{\pi}\int\limits_{\w_{th}}^\infty \!
\frac{d\w}{\w}\,\De\si_{1n}^{(n)}, \eeq
 \item[(iii)] the consistency conditions (valid to leading order in the
coupling $g$): \beq \eqlab{ncons} 0 =
\frac{1}{\pi}\int\limits_{\w_{th}}^\infty \!
\frac{d\w}{\w}\,\De\si_{1n}^{(p)} ,\,\,\,\, 0  =
\frac{1}{\pi}\int\limits_{\w_{th}}^\infty \!
\frac{d\w}{\w}\,\De\si_{1p}^{(n)}. \eeq
\end{subequations}
\end{itemize}

The first derivatives of the cross-sections that enter in
\Eqref{newsrs}, to leading order in $g$, arise through the
interference of Born graphs \Figref{Born_chpt}(a) with the graphs
in \Figref{Born_chpt}(b) and we find:
\begin{subequations}
\eqlab{piprod1}
\bea
\De\si_{1p}^{(p)}\equiv \De\si_{1p}^{(\pi^0 p)}+ \De\si_{1p}^{(\pi^+ n)}&=&
\frac{\pi C}{M^2x^2}\left\{ 2x\la [4+(1 -2\al) (2+\bar s+2x)]
+2 \bar s\la (\al+2) \right.\nn\\
&-& \left. \mu^2x \ln\frac{\be+\la}{\be-\la}+(2\al \bar s+1-x)
 \ln\frac{\al+\la}{\al-\la} \right\},\\
\De\si_{1n}^{(n)}\equiv\De\si_{1n}^{(\pi^0 n)}+ \De\si_{1n}^{(\pi^- p)}&=&
\frac{\pi C}{M^2x}\left\{ 2\la (2+2x-\bar s)
+\mu^2\ln\frac{\be+\la}{\be-\la} -\ln\frac{\al+\la}{\al-\la}
\right\}, \qquad\\
\De\si_{1n}^{(p)}\equiv \De\si_{1n}^{(\pi^0 p)}+ \De\si_{1n}^{(\pi^+ n)}&=&
\frac{2\pi C}{M^2x^2}\left\{\ln\frac{\al+\la}{\al-\la} +2\la (x\be -\bar s \al)
\right\},\\
\De\si_{1p}^{(n)}\equiv\De\si_{1p}^{(\pi^0 n)}+ \De\si_{1p}^{(\pi^- p)}&=&
\frac{2\pi C}{M^2x^2}\left\{(2\bar s\al-x) \, \ln\frac{\al+\la}{\al-\la} +2\la
(x -2\bar s ) \right\}.
\eea
\end{subequations}

Using the latter two expressions we easily verify the consistency
conditions given in \Eqref{ncons} \footnote{We checked that these
conditions are verified as well for the case of pseudoscalar $\pi
NN$ coupling.} while, employing the linearized SRs, we obtain:
\begin{subequations}
\eqlab{amms}
\bea
\kappa_p^{\rm (loop)} &=& \frac{M^2}{\pi e^2} \int\limits_{\w_{\mathrm{th}}}^{\infty}\!
\frac{d\w}{\w} \De\si_{1p}^{(p)}\nn\\
&=&\frac{g^2}{(4\pi)^2 } \left\{1 -
  \frac{\mu \,\left( 4 - 11{\mu }^2 + 3{\mu }^4 \right) }{\sqrt{1 - \quarter
{\mu }^2}}
  \arccos \frac{\mu }{2} - 6{\mu }^2+
  2{\mu }^2\left( -5 + 3\,{\mu }^2 \right) \ln \mu \right\}, \\
\kappa_n^{\rm (loop)} &=&\frac{M^2}{\pi e^2}
\int\limits_{\w_{\mathrm{th}}}^{\infty}\! \frac{d\w}{\w}
\De\si_{1n}^{( n)} =\frac{-2g^2}{(4\pi)^2 } \left\{2 -
\frac{\mu\,(2-\mu^2)}{\sqrt{1 - \quarter {\mu }^2}} \arccos
\frac{\mu }{2} - 2\mu^2 \ln \mu \right\}. \label{eq:bh}\eea
\end{subequations}
Exactly the same result is found in the case of pseudoscalar
pion-nucleon coupling and indeed \Eqref{amms} agrees with the
long-known one-loop calculation done by using standard
techniques~\cite{BeH}.  It is also worth noting that the result
given in \Eqref{amms} is not entirely in agreement with the chiral
perturbation theory calculation of Ref.~\cite{KuM}. The
discrepancy is apparently due to the fact that the ''infrared
regularized'' loop amplitudes exploited in~\cite{KuM} do not
satisfy the usual dispersion relations. Their analytic structure
in the energy plane is somewhat more complicated since there are
additional cuts due to explicit dependence on $\sqrt{s}$,
cf.~\cite{BeL}.

Finally, we would like to make an observation concerning the
chiral behavior of the one-loop result for the nucleon \amm.
Expanding \Eqref{amms} around the chiral limit ($m_\pi=0$), which
incidentally corresponds here with the heavy-baryon expansion, we have
\bea
\eqlab{expand_amms}
\kappa_p^{\rm (loop)} &=& \frac{g^2}{(4\pi)^2 }\left\{1 -2\pi
\mu-2\,(2+5\ln\mu)\,\mu^2+\frac{21\pi}{4}\,\mu^3
+ O(\mu^4)\right\}, \\
\kappa_n^{\rm (loop)} &=&\frac{g^2}{(4\pi)^2 }
\left\{-4 +2\pi \mu-2\,(1-2\ln\mu)\,\mu^2-\frac{3\pi}{4}\,\mu^3+ O(\mu^4)
\right\}.
\eea
The term linear in pion mass (recall that $\mu=m_\pi/M$) is the well-known
leading nonanalytic (LNA) correction.
On the other hand, expanding the same expressions around the large $m_\pi$ limit
we find
\bea
\kappa_p^{\rm (loop)} &=& \frac{g^2}{(4\pi)^2 }  \,(5-4\ln\mu)\frac{1}{\mu^2} 
+ O(\mu^{-4}), \\
\kappa_n^{\rm (loop)} &=&\frac{g^2}{(4\pi)^2 } \,2 (3-4\ln\mu)\frac{1}{\mu^2} +
O(\mu^{-4}).
\eea 
What is intriguing here is that the one-loop
correction to the nucleon \amm\ for heavy quarks  behaves as
$1/m_{quark}$ (where $m_{quark} \sim m_\pi^2)$, precisely as expected from a
constituent quark-model picture.  Here this is a result of subtle cancellations
in \Eqref{amms} taking place for large values of $m_\pi$. In contrast, the 
infrared regularization procedure~\cite{KuM} gives
the result which exhibits pathological
behavior with increasing pion mass and diverges for $m_\pi =2M$.

Since the expressions in \Eqref{amms} have the
correct large $m_\pi$ behavior they should be better suited for the
chiral extrapolations of the lattice results than the usual
heavy-baryon expansions or the ``infrared-regularized'' relativistic theory. 
This point is clearly demonstrated by
\Figref{chibehavior}, where we plot the $m_\pi$-dependence of
the full [\Eqref{amms}],  heavy-baryon, and infrared-regularization~\cite{KuM}
leading order result for the magnetic moment of the proton and the neutron, 
in comparison  to recent lattice data~\cite{Zan04}.  In presenting these results 
we have added a constant shift (counter-term $\kappa_0$) to the magnetic
moment, i.e.,
\bea
\mu_p&=&(1+\kappa_{0p}+\kappa_p^{\rm (loop)})(e/2M),\\
\mu_n&=&(\kappa_{0n}+\kappa_n^{\rm (loop)}) (e/2M)
\eea
 and fitted it to the known experimental value of the magnetic
moment at the physical pion mass, $\mu_p\simeq 2.793$ and $\mu_n\simeq -1.913$, shown
by the open diamonds in the figure. For the value
of the $\pi NN$ coupling constant we have used $g^2/4\pi = 13.5$. The $m_\pi$-dependence
away off the physical point is then a prediction of the theory. The figure clearly
shows that the SR results, shown by the dotted lines,
is in a better agreement with the behavior obtained in lattice gauge
simulations.

 It is therefore convenient to use the SR results for the parametrization
of lattice data. For example, we consider the following two-parameter form:
\begin{subequations}
\eqlab{paran}
\bea
\mu_p&=&\left(1+\frac{\tilde \kappa_{0p}}{1+a_p m_\pi^2}+\kappa_p^{\rm (loop)}\right)\frac{e}{2M},\\
\mu_n&=&\left(\frac{\tilde\kappa_{0n}}{1+a_n m_\pi^2}+\kappa_n^{\rm (loop)}\right)\frac{e}{2M}\,,
\eea
\end{subequations}
where $\tilde \kappa_{0p}$ and $\tilde \kappa_{0n}$ are fixed to 
reproduce the experimental magnetic moments
at the physical $m_\pi$. The parameter $a$ can be fitted to lattice data.
The solid curves in \Figref{chibehavior} represent the result of such a single parameter
fit to the lattice data of Ref.~\cite{Zan04}
for the proton and neutron respectively, where $a_p=1.6/M^2$ and $a_n=1.05/M^2$,
$M$ is the physical nucleon mass.

In conclusion, we have presented a new sum rule which in essence
can be viewed as the first derivative of the well-known GDH sum
rule w.r.t.\ the anomalous magnetic moment. The attractive feature
of this new sum rule is that it established a {\it linear}
relation between the \amm\ and the cross section (in contrast to 
the GDH SR where the relation is quadratic), allowing an evaluation of
loop corrections to the \amm\ by computing a total cross section
of a corresponding photoabsorption process to one loop {\it lower}
than the desired result and then integrating it over energy.  As
an example, we reproduced in this way the celebrated Schwinger
correction to the electron \amm, as well as considered the one
pion-nucleon loop correction to the nucleon magnetic moment.

Of course, the results presented herein are not the end of such
applications.  Indeed, one can envision extension both to higher
order calculations by use of one loop inputs and/or application to
other sum rules, such as those for polarizabilities. However, we leave
these as challenges for future work.

\begin{acknowledgments}
This work is supported in part by DOE grant no.\
DE-FG02-04ER41302 and contract DE-AC05-84ER-40150 under
which the Southeastern Universities Research Association (SURA)
operates the Thomas Jefferson National Accelerator Facility.  
The work of BRH is also supported in part by
NSF PHY-02-44801 and he would like to thank the JLab theory group
for hospitality while this work was in progress.
\end{acknowledgments}

\newpage

\begin{figure}[h,b,t,p]
\centerline{
  \epsfxsize=7cm
  \epsffile{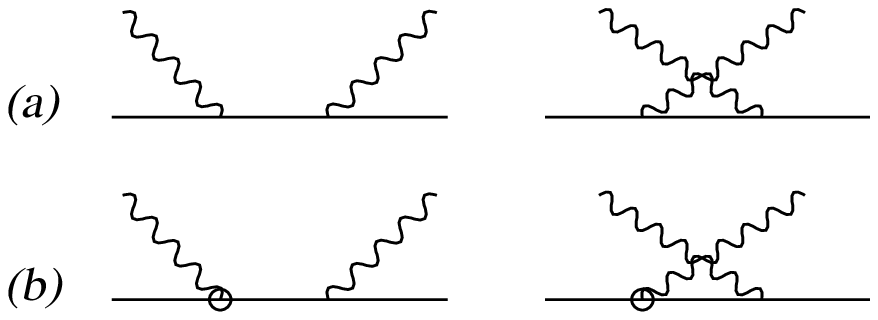}
}
\caption{Tree-level Compton scattering graphs. The circled vertex corresponds to the
Pauli coupling.}
\figlab{compton_qed}
\end{figure}

\begin{figure}[h,b,t,p]
\centerline{
  \epsfxsize=8cm
  \epsffile{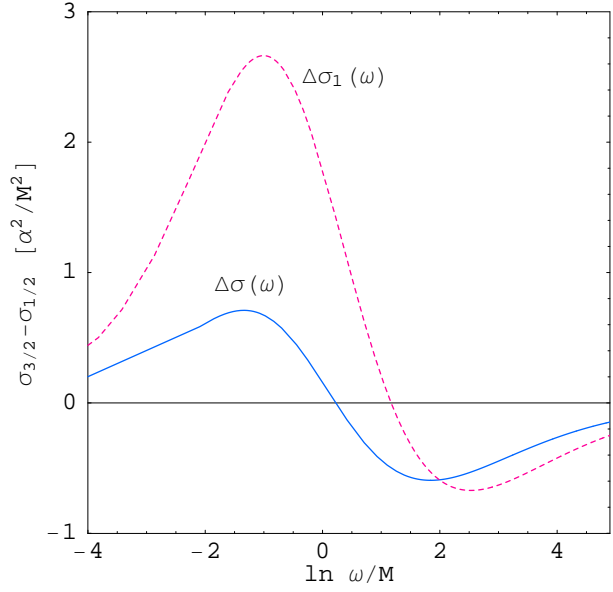}
}
\caption{GDH integrands for Compton scattering cross-sections (in units of
$\alpha^2 / M^2$), corresponding to
Eqs.~\eref{gdh} and
\eref{gdh2}.}
\figlab{qedfig}
\end{figure}

 \begin{figure}[h,b,t,p]
\centerline{
  \epsfxsize=12cm
  \epsffile{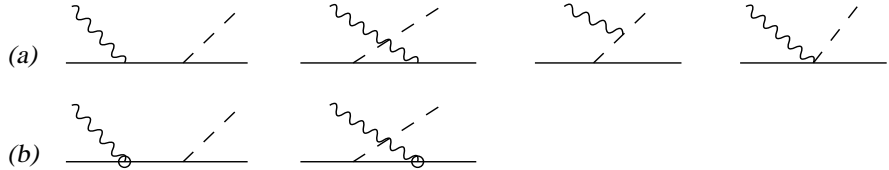}
}
\caption{Tree-level pion photoproduction graphs. The circled vertex corresponds to
the Pauli coupling.}
\figlab{Born_chpt}
\end{figure}

\begin{figure}[h,b,t,p]
\centerline{
  \epsfxsize=8cm
  \epsffile{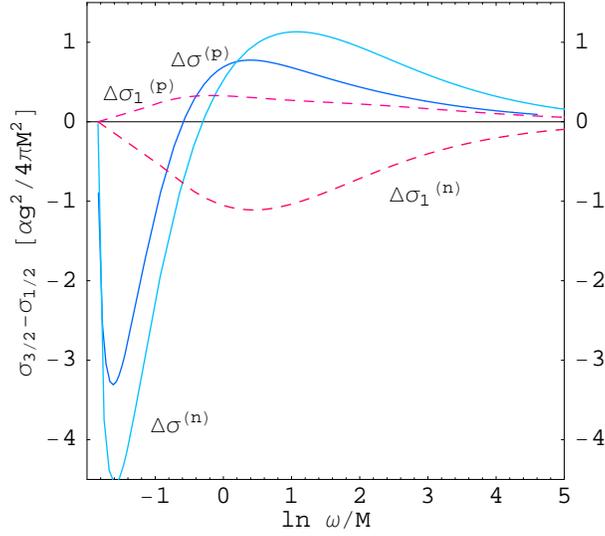}
}
\caption{GDH integrands (in units of $C/ M^2$) for Born-level pion
photoproduction cross sections,
corresponding to
Eqs.~\eref{piprod} and \eref{piprod1}.}
\figlab{chpt1}
\end{figure}

\begin{figure}[h,b,t,p]
\centerline{
  \epsfxsize=8cm
  \epsffile{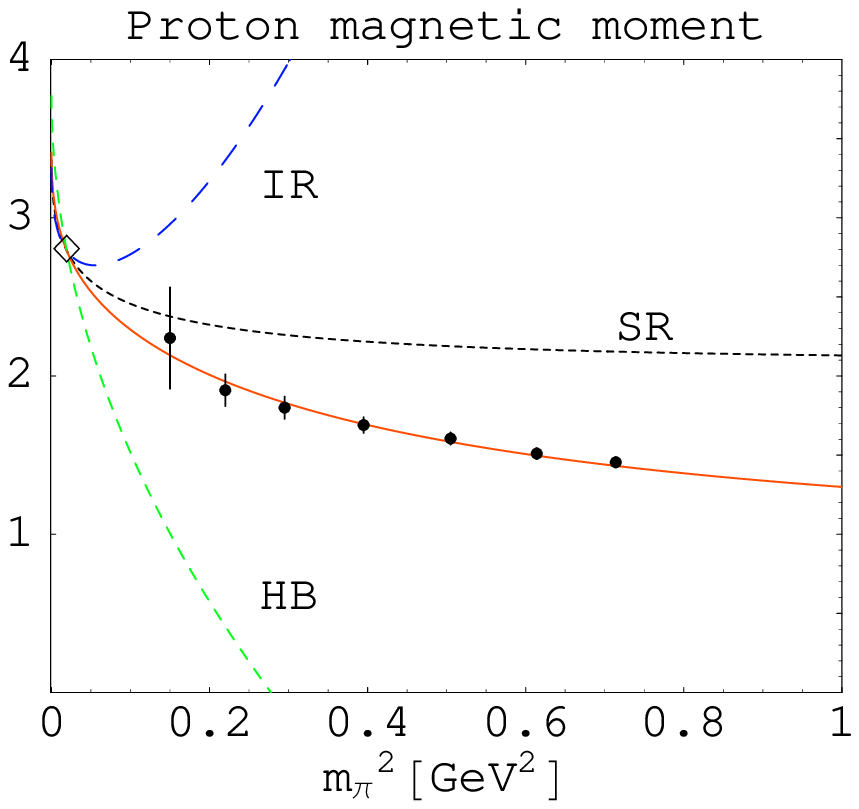}
\epsfxsize=8cm  \epsffile{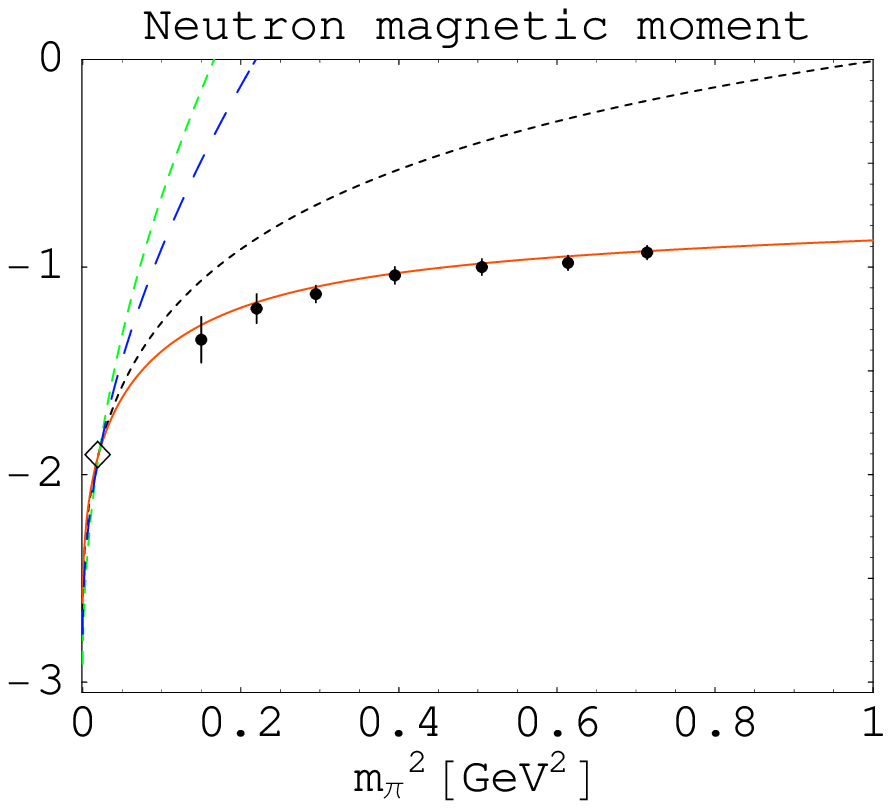}
}
\caption{Chiral behavior of the proton and neutron magnetic moments (in nucleon magnetons) 
to one loop
compared with lattice data. ``SR'' (dotted lines) represents
the full result given
by Eqs.~\eref{amms}, ``IR'' (blue long-dashed lines) the infrared-regularized relativistic
result, ``HB'' (green dashed lines) the LNA term in the heavy-baryon
expansion \Eqref{expand_amms}. Red solid lines are the fit of the parametrization 
in \Eqref{paran} based on the SR result. Data points are results of 
lattice simulations~\cite{Zan04}. The open diamonds represent the experimental values
at the physical pion mass.}
\figlab{chibehavior}
\end{figure}

\end{document}